# False Summit and Silent Drift: A Failure Taxonomy and Efficiency Analysis of LLM-Assisted Multiphysics Simulation in an Open-Source Framework

Chi-Chun Chiang[a], Shih-Ming He[b,*]

[a] *Interdisciplinary Program of Electrical Engineering and Computer Science, Chung Yuan Christian University, Taoyuan City 320314, Taiwan*

[b] *Department of Electronic Engineering, Chung Yuan Christian University, Taoyuan City 320314, Taiwan*

* *Corresponding author: shihming.he@gmail.com*

## Abstract

Multiphysics simulation of semiconductor processes requires simultaneous command of governing equations, platform syntax, numerical methods, and process physics. This combination raises a steep entry barrier for researchers without formal training in computational science. We test whether large language model (LLM) assistance, staged model development, and curated domain literature can lower this barrier, and we characterize the failure modes that persist when they do. In a fully documented case study, a researcher with no prior experience in Multiphysics Object-Oriented Simulation Environment (MOOSE) or chemical vapor deposition (CVD) developed a six-stage, fully coupled low-pressure CVD (LPCVD) model for graphene growth. We compared two assistants, GPT (via Codex) and Claude (via Claude Code), under three conditions: direct prompting, staged development without literature, and staged development with curated literature. In this case study, direct prompting never produced a converging simulation. Staged development succeeded but exposed two reproducible failure classes, which we term false summit and silent drift. A false summit is a simulation that converges and appears physically plausible while implementing incorrect physics. Two instances dominate the dataset: a non-conservative advection kernel applied to weakly compressible flow (FM1; $\Delta T \approx 970$ K, $\nabla\cdot\mathbf{u} \approx 10^2$ $s^{-1}$) produced spurious methane depletion that, in its diffusion-dominated limit, reduced the $CH_4$ mass fraction to ~27% of the inlet value (~73% depletion) yet remained morphologically indistinguishable from genuine surface chemistry, and a chain of surface-reaction formulation errors (FM2), including an omitted $(1-\theta_g)$ self-limiting term and an inverted gas-density expression, drove surface coverage to saturation roughly 200 times faster than the physically expected rate. In the FM2 case, the AI agent justified successive incorrect results with Damköhler-number reasoning across three model revisions, rather than identifying the underlying defects. Silent drift is the undetected propagation of an unverified parameter or geometric assumption through successive development stages; in one case, an inlet-velocity error survived 13 prompts before detection. Curated-literature directives reduced Stage 6 prompt counts by 91% for GPT and 64% for Claude, shifting the dominant effort from reactive debugging to constructive physics formulation; neither failure class, however, disappeared. We provide an eight-category failure taxonomy, model-specific human-oversight strategies, and a stage-gated workflow that makes both failure classes detectable at stage boundaries.

## Nomenclature

| Symbol | Definition (units) |
|---|---|
| $c_p$ | Specific heat capacity (J $kg^{-1}$ $K^{-1}$) |
| $D_i$ | Diffusivity of species *i* ($m^2$ $s^{-1}$) |
| $k$ | Thermal conductivity (W $m^{-1}$ $K^{-1}$) |
| $T$ | Gas temperature (K) |
| $T_s$ | Surface temperature (K) |
| $\boldsymbol{u}$ | Velocity field (m $s^{-1}$) |
| $Y_i$ | Mass fraction of species *i* |
| $A_{dep}$, $A_{etch}$ | Pre-exponential factors for deposition / etching |
| $E_{dep}$, $E_{etch}$ | Activation energies for deposition / etching (kJ $mol^{-1}$) |
| $J_{dep}$, $J_{etch}$ | Deposition / etching molar fluxes (mol $m^{-2}$ $s^{-1}$) |
| $k_{dep}$, $k_{etch}$ | Arrhenius rate constants for deposition / etching |
| $p_{CH4}$ | Partial pressure of $CH_4$ (Pa) |
| $p_{H2}$ | Partial pressure of $H_2$ (Pa) |
| $R$ | Universal gas constant (J $mol^{-1}$ $K^{-1}$) |
| $R_{dep}$, $R_{etch}$ | Deposition / etching rates in the coverage ordinary differential equation (ODE) ($s^{-1}$) |
| $Da$ | Damköhler number |
| $Kn$ | Knudsen number |
| $Re$ | Reynolds number |
| $n$ | Reaction order for $CH_4$ partial pressure |
| $\alpha$ | Exponent of the self-limiting coverage factor |
| $\alpha_{dep}$ | Sticking / deposition coefficient in the concentration form |
| $\beta$ | Reaction order for $H_2$ partial pressure |
| $\Gamma_{mono}$ | Monolayer site density (mol $m^{-2}$) |
| $\theta_g$ | Graphene surface coverage |
| $\theta_{ss}$ | Steady-state surface coverage |
| $\rho$ | Gas density (kg $m^{-3}$) |

# 1. Introduction

## 1.1 The entry barrier and two classes of invisible failure

Open-source finite element platforms such as Multiphysics Object-Oriented Simulation Environment (MOOSE) enable publication-quality simulation of coupled thermal, fluid, and reactive systems without licensing costs, making them increasingly attractive to researchers entering semiconductor process engineering[1]. The barrier to adoption is not access to software. It is the compound expertise required to use such platforms correctly: a newcomer building a coupled heat-flow-species-reaction model must navigate governing equation selection, discretization schemes, solver settings, and boundary condition types simultaneously, and each choice can produce a simulation that runs without error yet implements incorrect physics. Recognizing incorrect physics requires exactly the domain fluency the newcomer is trying to develop. This self-referential difficulty produces the first failure class studied here, which we call the false summit: a simulation that appears correct below the operator's knowledge threshold while concealing errors visible only above it. The second class, silent drift, is procedural rather than physical: an unverified numerical parameter or geometric assumption introduced at an early stage propagates quietly through later stages because nothing in the development loop forces it to be re-examined. In the case study reported below, an inlet velocity specified at roughly four times its correct value passed the stage at which it was introduced and survived thirteen further prompts before detection. Neither failure class produces a solver error, and both become harder to detect as physics coupling deepens.

Interest in large language model (LLM)-assisted scientific code generation has motivated several frameworks for solver automation[2]. OpenFOAMGPT and MetaOpenFOAM show that LLM agents can automate OpenFOAM workflows for incompressible and single-physics problems with pass rates of 80–85%[3,4], and MooseAgent applies an analogous multi-agent approach to MOOSE, reaching 93% success on heat transfer and mechanics benchmarks[5]. FEABench evaluates LLM performance for multiphysics finite element reasoning in COMSOL and finds that performance degrades substantially on coupled problems[6]. These results are encouraging, but the problems evaluated are predominantly single-physics, and the errors involved are largely syntactic: they surface as convergence failures or malformed inputs. Hallucination taxonomies for code generation likewise categorize failures into task-requirement, factual-knowledge, and context conflicts, all of which appear in the present dataset[7,8]. To our knowledge, existing taxonomies do not emphasize the failure mode in which an agent generates physically sophisticated, internally consistent wrong arguments in defense of an incorrect result, so that further prompting deepens the justification rather than correcting the physics. Documenting that mode, and the workflow conditions under which it arises, is the purpose of this study. We identify two reproducible failure modes: FM1 (false physical depletion caused by non-conservative transport) and FM2 (confident misvalidation, where physically plausible interpretations conceal incorrect governing equations).

### 1.2 Why LPCVD graphene growth is a demanding test

Graphene growth by LPCVD on copper foil was selected as the application case because it couples heat transfer, weakly compressible gas-phase transport, heterogeneous surface chemistry, and self-limiting growth kinetics, all of which are well documented in the literature and therefore admit ground-truth physical validation. The reactor and operating conditions used here are taken from our previous experimental graphene-CVD study[9], which provides an experimental basis for the simulated regime rather than a purely literature-derived one[10–12]. The 970 K axial thermal gradient produces an approximately fourfold gas density variation, so the condition $\nabla\cdot\mathbf{u} \neq 0$ is both unavoidable and consequential. Competing deposition and hydrogen-etching pathways create a surface reaction system in which the self-limiting coverage factor $(1-\theta_g)$ is physically essential yet easy to omit[13]. Both principal failure modes documented here arise from this specific physics rather than from incidental implementation choices, and the need for artificial intelligence (AI) integration in CVD simulation has been highlighted[12]. The present study is a direct empirical response.

### 1.3 Contributions

The study makes four contributions. First, it shows empirically that, in this LPCVD graphene case, direct naive prompting failed in every attempt on a fully coupled multiphysics problem, suggesting that staged physics development is practically necessary for similarly complex cases and placing it on an evidential rather than purely stylistic footing. Second, it characterizes two reproducible failure modes, FM1 (non-conservative species transport) and FM2 (confident misvalidation), that are invisible to newcomers without domain expertise and persist under literature-guided conditions, extending existing LLM failure catalogs into the physics-semantics domain[7,8]. Third, it quantifies the effect of curated literature directives: Stage 6 prompt counts fell by 91% for GPT and 64% for Claude, and the dominant effort category shifted from debugging to equation formulation. Fourth, it distills these findings into a stage-gated workflow (Section 6) that makes false summits and silent drift detectable at stage boundaries rather than after they have contaminated the coupled model.

## 2. Methods

### 2.1 Researcher baseline

The lead researcher had an undergraduate background in electrical engineering, with basic Python and C programming experience, and no formal coursework in fluid mechanics, heat transfer, or transport phenomena, and no prior exposure to MOOSE, finite element simulation, or CVD process modeling. This baseline represents a realistic entry point for newcomers: enough general engineering and programming background to operate the tools and read the governing equations at a conceptual level, but without the domain-specific fluency in transport physics, the platform-specific expertise in MOOSE, or the process-specific knowledge of CVD needed to independently diagnose the failure modes encountered. Because this baseline precludes diagnosing physical errors by expert intuition, every validation criterion in this study was defined as an external, independently computable reference rather than a judgment of physical plausibility. Stage criteria were checked against hand calculations, analytical limits, conservation balances, and published kinetic parameters (Section 2.4, Table 2), and AI-generated explanations were never accepted as evidence of correctness. Documenting the baseline, therefore, matters methodologically: it defines the knowledge threshold below which both FM1 and FM2 are undetectable without such external validation, and it is precisely this threshold that the staged, criterion-gated workflow is designed to compensate for.

### 2.2 Simulation target, governing equations, and minimum physics

The simulation target was LPCVD graphene growth on copper foil in a horizontal tube furnace. All simulations used a 2D Cartesian geometry representing the axial–vertical cross-section of the reactor. A rotationally symmetric ($R_Z$) formulation is structurally inadequate because the off-axis copper substrate is not symmetric about the tube centerline; initial attempts using $R_Z$ geometry forced a complete architectural rebuild at Stage 6. Reactor parameters are summarized in **Table 1**.

**Table 1.** Reactor geometry, operating conditions, and derived physical parameters used in all simulations.

| Parameter | Value | Source |
|---|---|---|
| Tube length | 60 cm | Experimental configuration adapted from previous work[9] |
| Tube inner diameter | 2.54 cm | |
| Substrate dimensions (length / thickness) | 2 cm / 30 μm | |
| Furnace temperature (heated zone, 10–50 cm) | 1273 K | |
| Operating pressure | 800 mTorr (106.6 Pa) | |
| Inlet flow rates (Ar / $H_2$ / $CH_4$) | 80 / 20 / 0.5 sccm | |

| Parameter | Value | Source |
| --- | --- | --- |
| Inlet velocity (calculated) | 3.45 m/s | Calculated from reactor geometry and operating conditions |
| $CH_4$ diffusivity at 1273 K | ~ 0.24 $m^2/s$ | Estimated based on literature[14] |
| Reynolds number / Knudsen number | ≈ 45 / ≈ 0.009 | Derived from reactor operating conditions |
| Kinetic pre-exponential factors | Literature estimates; not experimentally calibrated | Derived from literature[13] |

A physically valid model must represent four coupled domains. Heat transfer: the 970 K gradient between the inlet (300 K) and the heated furnace zone (1273 K) produces an approximately fourfold change in gas density and drives $\nabla\cdot\mathbf{u} \approx 10^2\ s^{-1}$, the root cause of FM1 when incompressible advection is used for species transport. Weakly compressible flow: the system is laminar (Re ≈ 45) and in the continuum regime (Kn ≈ 0.009), with a characteristic double-peak velocity profile created by the off-axis substrate[15,16]. Multi-species transport: $CH_4$ diffusivity at 1273 K is approximately 0.24 $m^2/s$, so the concentration field is nearly uniform in the absence of surface reaction; any pre-reaction concentration gradient is therefore a diagnostic signal of non-physical transport[14]. Surface reaction with self-limiting growth: the competing deposition and $H_2$ etching pathways require explicit coverage tracking through a bounded ordinary differential equation[13]. More generally, FM2 arose when physically plausible interpretations repeatedly defended models containing unresolved formulation, coupling, or parameter errors despite successful numerical convergence.

The governing equations for the fully coupled Stage 6 model are as follows.

Energy conservation:

$$\rho c_p \left(\frac{\partial T}{\partial t} + \boldsymbol{u}\cdot\nabla T\right) = \nabla\cdot(k\nabla T)$$

Conservative species transport (required to avoid FM1):

$$\frac{\partial(\rho Y_i)}{\partial t} + \nabla\cdot(\rho \boldsymbol{u} Y_i) = \nabla\cdot(\rho D_i \nabla Y_i)$$

Using $\nabla\cdot(\mathbf{u}Y_i)$ instead of $\nabla\cdot(\rho\mathbf{u}Y_i)$, the incompressible form, is the root cause of FM1 under weakly compressible conditions.

Surface coverage evolution:

$$\frac{d\theta_g}{dt} = \frac{J_{\text{dep}} - J_{\text{etch}}}{\Gamma_{\text{mono}}}$$

where

$$J_{\text{dep}} = k_{\text{dep}}(T_s)\, p_{\text{CH}_4}^{n}\left(1-\theta_g\right)^{\alpha}$$

$$J_{\text{etch}} = k_{\text{etch}}(T_s)\, p_{\text{H}_2}^{\beta}\theta_g$$

Omission of the self-limiting factor $(1-\theta_g)^{\alpha}$ represents one example of the formulation errors encountered during FM2, where $\rho$ is the gas density, $c_p$ the specific heat capacity, $k$ the thermal conductivity, $T$ the temperature, $\boldsymbol{u}$ the velocity field, $Y_i$ and $D_i$ the mass fraction and diffusivity of species $i$, $\theta_g$ the graphene surface coverage, $J_{dep}$ and $J_{etch}$ the deposition and etching fluxes, $\Gamma_{mono}$ the monolayer site density, $k_{dep}$ and $k_{etch}$ the Arrhenius rate constants ($k_{dep} = A_{dep} \exp(-E_{dep}/RT_s)$, in which $A_{dep}$ and $A_{etch}$ are the pre-exponential factors, $E_{dep}$ and $E_{etch}$ the activation energies, and $R$ the universal gas constant), $T_s$ the surface temperature, $p_{CH_4}$ and $p_{H_2}$ the partial pressures of $CH_4$ and $H_2$, and $n$, $\alpha$, and $\beta$ the reaction orders. The kinetic parameters used for each model variant are listed in **Table 6**.

## 2.3 Platform: MOOSE and the rationale for open source

MOOSE supports arbitrary physics coupling through a kernel-based finite element architecture in which each physical term corresponds to a distinct code object whose parameters and valid usage contexts are defined in the application programming interface (API) documentation.[1] This architecture proved central to resolving both failure modes. Fixing FM1 required establishing that the available incompressible advection kernel does not accept a density parameter and re-implementing species transport within the weakly compressible Navier–Stokes (WCNS) finite-volume framework, with ρY as the conserved quantity. Fixing FM2 required auditing the surface reaction auxiliary kernel term by term against published rate expressions. Both operations were enabled by direct access to source code and API documentation, an epistemic transparency that open-source platforms provide without vendor mediation and that gives them a clear practical advantage for LLM-assisted debugging in this study, even though similar insight might in principle be obtained from closed commercial solvers with sufficient documentation and support.

## 2.4 Six-stage development ladder

The simulation was constructed in six stages of strictly increasing physical complexity (**Table 2**). A quantitative validity criterion was defined before each stage began, and advancement required independent verification of that criterion, not merely numerical convergence. This pre-commitment is the structural mechanism that makes false summits detectable at stage boundaries rather than allowing them to accumulate silently in the fully coupled model.

**Table 2.** Development stages and quantitative validation criteria used throughout model construction.

| Stage | Physics | Validity criterion |
|---|---|---|
| Stage 1: Temperature field | Heat conduction and furnace-zone profile | T = 300 K at inlet and outlet; T = 1273 K in heated zone (10–50 cm) |
| Stage 2: Species transport | $CH_4$ transport and diffusion coupling (passive) | $\Delta Y_{CH4} \approx 0$ without surface reaction; no spurious depletion gradient anywhere in the domain |
| Stage 3: Laminar flow field | Velocity profile from inlet flow conditions | Centerline velocity = 2× inlet velocity (Poiseuille profile); error < 1% |
| Stage 4: Heat–flow coupling | Thermal expansion and coupled transport | Velocity increase in the heated zone is consistent with density reduction under ideal-gas expansion |
| Stage 5: Weakly compressible transport | Conservative species transport (no reaction) | Mass conservation error < 1%; zero-reaction condition yields $\Delta Y_{CH4} \approx 0$ |
| Stage 6: Surface reaction and coverage evolution | Graphene growth and $H_2$ etching kinetics | $0 \le \theta_g \le 1$ everywhere; inlet–outlet $CH_4$ flux balance < 0.3%; zero-reaction recovers Stage 5 |

## 2.5 Prompting conditions and failure classification

Three prompting conditions were evaluated. Condition 0 (direct prompting) consisted of a single prompt requesting a complete LPCVD graphene simulation, with no staged structure and no literature. Condition 1 (staged, without literature) used the six-stage ladder with geometry and process parameters only. Condition 2 (staged, with literature) used the same ladder augmented by four explicit physics directives: $H_2$ participates in both activation and etching; species transport must use $\nabla\cdot(\rho\mathbf{u}Y_i)$; surface coverage evolves through a bounded ordinary differential equation (ODE); and partial pressure governs nucleation density. These constraints were stated as imperative directives within the generation scope rather than as background context, a formulation that consistently produced higher adherence and is consistent with the principle-specification approach used in retrieval-augmented computational fluid dynamics (CFD) agents[3].

The human operator classified failures into eight categories (**Table 3**) in real time, where a failure is any prompt cycle requiring human-initiated correction before stage advancement could proceed. The classification was applied consistently across all four model–condition combinations: GPT/without literature, GPT/with literature, Claude/without literature, and Claude/with literature. Development was conducted between April and May 2026 using GPT (GPT-5.4 and, from Stage 6 Prompt 14 of the without-literature condition, GPT-5.5) via Codex and Claude (Opus 4.6) via Claude Code; the GPT version transition is noted as a possible confound in the Limitations.

**Table 3.** Eight-category failure taxonomy with representative examples.

| Code | Category | Description | Representative instance |
|---|---|---|---|
| FC1 | Geometry / coordinate | Axis inversion, $R_Z$ vs. Cartesian, substrate omission | GPT: x = radial, y = axial inverted (Stages 1–3); Cu substrate omitted from Stage 4 |
| FC2 | Missing physics | Absent advection kernel, etching flux, or self-limiting term | GPT: no $CH_4$ advection in initial Stage 6; Claude: missing $(1-\theta_g)$ in FM2 |
| FC3 | Kernel API mismatch | Kernel rejects required parameter; wrong kernel class selected | Claude: INSFVScalarFieldAdvection rejects ρ parameter |
| FC4 | Equation form | Non-conservative form in compressible regime; Boussinesq misuse | Both models, FM1: $\nabla\cdot(\mathbf{u}Y)$ instead of $\nabla\cdot(\rho\mathbf{u}Y)$ |
| FC5 | Unit / parameter error | Molar mass in g/mol used as kg/mol; fourfold velocity overestimate; 53× $A_{dep}$ error | GPT: inlet velocity 14.6 vs. 3.45 m/s; Claude: $M_C$ = 12 as kg/mol |
| FC6 | Numerical stability | Face/cell flux mismatch; time-step overshoot; $\theta_g > 1$ from explicit integration | Claude: cell-centered $\rho\mathbf{u}$ vs. face Rhie–Chow flux (4.4% error) |
| FC7 | Module coupling | Operator splitting decouples fields; hardcoded variables sever coupling | Claude: GrapheneCoverageAux hardcoded inlet $Y_{CH4}$ |
| FC8 | Confident misvalidation | Physically coherent reasoning defending an incorrect result | Claude, FM2: Damköhler argument for missing $(1-\theta_g)$ term (three prompts) |

## 3. Results

### 3.1 Direct prompting fails universally

Given a single prompt requesting a complete LPCVD graphene simulation, both GPT and Claude produced MOOSE input files that failed to converge in every attempt. Observed failure types included incompatible combinations of physics modules, incorrect boundary condition variable types, missing coupling terms between physics blocks, and solver settings inappropriate for the low-pressure, weakly compressible regime. None of the resulting error messages offered guidance that a newcomer without domain expertise could act on. This outcome mirrors FEABench and MooseAgent, which document that single-physics automation works reliably while fully coupled problems expose the limits of zero-shot generation[5,6]. The practical implication in this case is clear: for problems of comparable coupling, staged physics development functioned as a structural necessity rather than a stylistic preference. A newcomer who begins with direct prompting must diagnose coupled failures across all physics simultaneously, which is precisely the situation in which the false summit problem is most severe and least detectable.

### 3.2 Staged workflow and interaction effort

The workflow structure was identical across conditions and models (**Figure 1**). For each stage, the operator composed a stage-level prompt specifying the physics scope, boundary conditions, and the predefined acceptance criterion; the LLM generated a complete MOOSE input file; the operator ran the simulation, evaluated the output against the validity criterion by an independent calculation, and either advanced or initiated a correction cycle. Stage closure was conditional on physical criterion satisfaction, never on convergence alone. One formulation detail proved consequential: literature constraints stated as explicit generation requirements, such as "species transport must use $\nabla\cdot(\rho \mathbf{u} Y_i)$", produced higher adherence than the same information presented as background context. Domain literature should therefore be reformulated as imperative directives before incorporation into prompts.

Two features of the interaction effort stand out (as shown in **Figure 2** and **Table 4**). First, Stage 6 dominates the conditions without literature: it accounts for 23 of 36 GPT prompts (64%), 11 of 16 Claude prompts (69%), and 73 of 103 GPT conversation turns. Second, literature guidance acts almost entirely on this Stage 6 bottleneck. With curated directives, Stage 6 fell to 2 prompts for GPT (a 91% reduction) and 4 for Claude (a 64% reduction), while Stages 1–5 costs were essentially unchanged. In this case, literature guidance did not broadly improve simulation performance across stages; it specifically removed the architecturally incorrect answers that made the surface-reaction and coverage stages expensive.

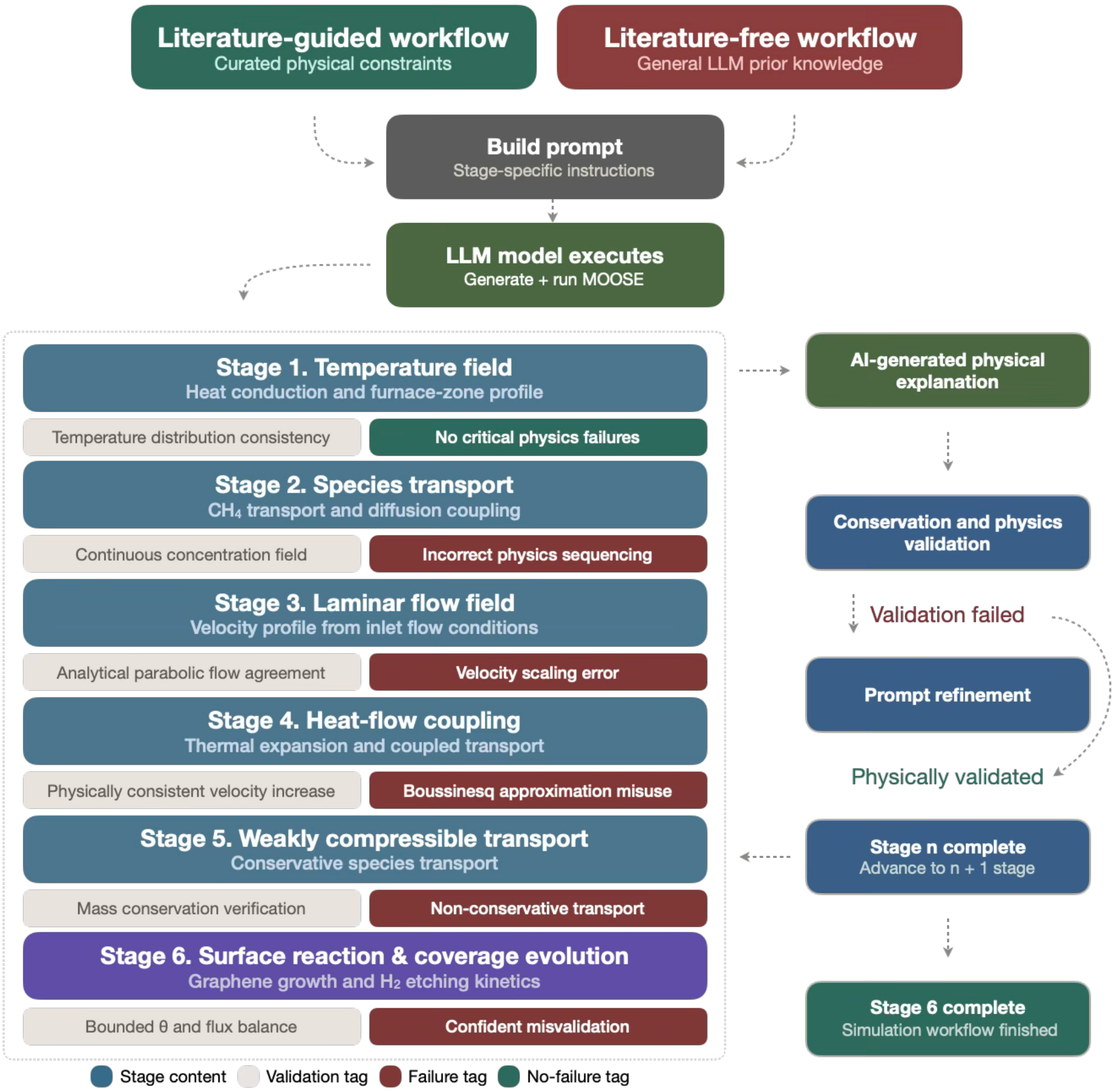

**Figure 1.** Failure discovery and validation workflow for LLM-assisted multiphysics simulation development in the MOOSE framework. Models generated at each stage are evaluated against predefined physical validation criteria before progression to the next stage.

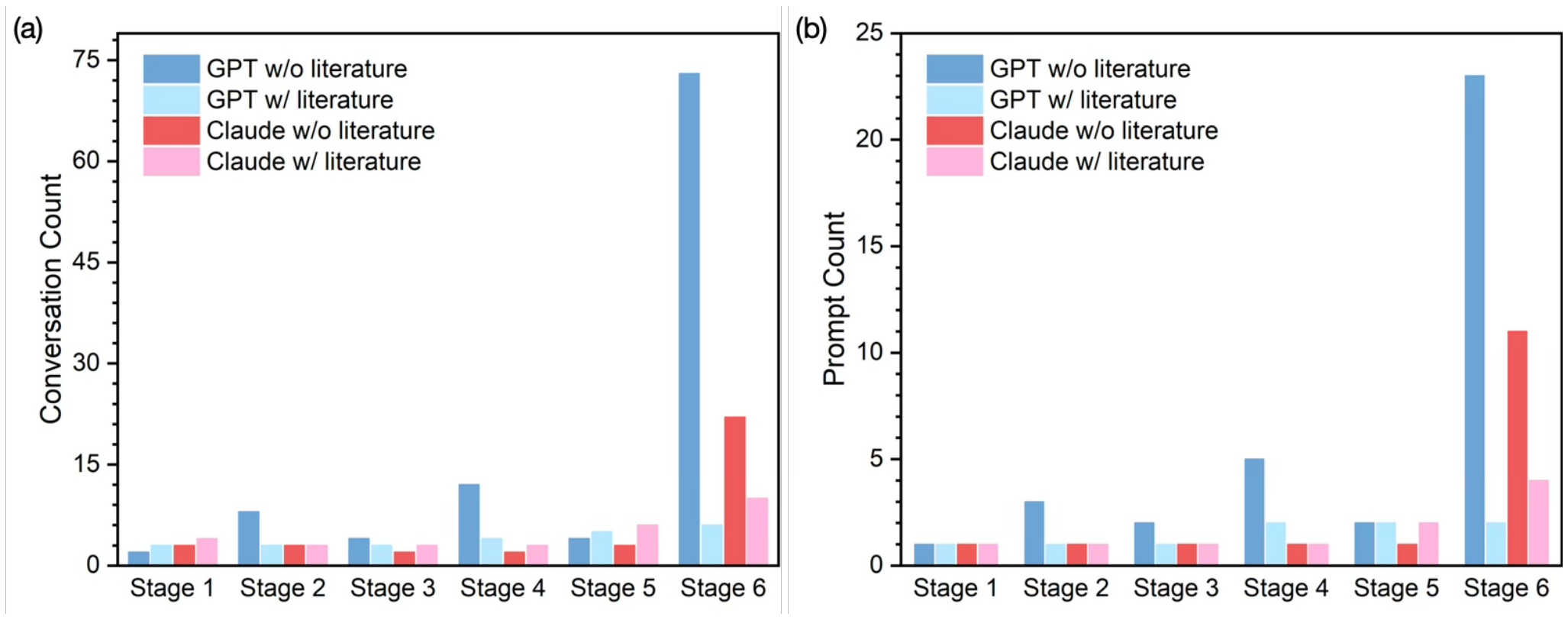

**Figure 2.** Effect of literature guidance on human–LLM interaction effort during staged model development. (a) Conversation counts and (b) prompt counts required to complete each development stage for GPT and Claude under literature-guided and literature-free workflows. Stage 6 exhibits the largest interaction burden and the greatest benefit from literature guidance.

**Table 4.** Prompt counts, conversation statistics, and development effort across workflow conditions.

| Metric | GPT w/o Literature | GPT w/ Literature | Claude w/o Literature | Claude w/ Literature | Reduction (GPT / Claude) |
|---|---|---|---|---|---|
| Total prompts (Stages 1–6) | 36 | 9 | 16 | 10 | 75% / 37.5% |
| Stages 1–5 prompts | 13 | 7 | 5 | 6 | — |
| Stage 6 prompts | 23 | 2 | 11 | 4 | 91% / 64% |
| Total conversation turns | 103 | ~ 24 | 35 | ~ 40–45 | 77% / ~ 14% increase† |
| Stage 6 turns | 73 | ~ 6 | 22 | ~ 10 | — |
| Major corrections (human-initiated) | 12 | 4–5 | 12 | 5–6 | ~ 58–67% / ~ 50–58% |
| First-pass convergence rate* | ~ 52% | 100% | ~ 85% | 100% | — |
| Effort, Stage 6# | ~ 270 min | ~ 62 min | ~ 17.7 hr | ~ 8–12 hr | ~ 77% / ~ 32–55% |
| Effort, all stages# | ~ 7 hr | ~ 4.2 hr | ~ 23–25 hr | ~ 16–22 hr | ~ 40% / ~ 12–36% |

*† Higher conversation counts under literature-guided conditions reflect longer verification exchanges per prompt.*
** First-pass convergence rate denotes the fraction of stage-level prompts whose generated input file converged on first execution, before any human-initiated correction.*
*# GPT effort represents active keyboard time, whereas Claude effort includes token-rate-limited waiting periods and is reported as wall-clock time; the two measures are therefore not directly comparable.*

## 3.3 Stages 1–5: early failure patterns and silent drift

Stages 1 through 5 under Condition 1 confirmed that both models handle single- and two-physics problems, but with qualitatively different error patterns that foreshadow Stage 6 behavior. GPT required 13 prompts for Stages 1–5, compared to Claude's 5. The largest single early-stage cost for GPT was Stage 4, which consumed five prompts: the first attempt applied the Boussinesq approximation, which is appropriate only for small density variations and therefore invalid for the 970 K gradient here[17,18], and three further prompts were needed to replace it with full variable-density weakly compressible Navier–Stokes. Claude implemented variable-density flow correctly on the first Stage 4 prompt. A geometrically consequential FC1 failure also occurred at GPT Stage 4: the copper substrate was omitted from the initial input file. The omission had no visible consequence in Stages 1–3, where the substrate plays no dynamically active role, and surfaced only when the coupled flow stage failed to produce the characteristic double-peak velocity profile. It is a delayed false summit embedded within the stage structure itself. The clearest instance of silent drift arose at GPT Stage 3. The inlet velocity was specified as 14.6 m/s rather than the correct 3.45 m/s (calculated from 100.5 sccm total flow at 800 mTorr, 300 K, and the 1-inch tube cross-section). The value passed the Stage 3 check because the criterion examined flow structure, not absolute magnitude, and the error was then propagated through Stages 3, 4, and 5 before being identified at Stage 6, Prompt 16, thirteen prompts after introduction. It is the longest-running silent error in the dataset and the clearest demonstration that numerical parameters must be verified by independent calculation rather than inferred from visual field plausibility, a principle consistent with the human-oversight requirements documented for LLM-assisted CFD agents.[3] This episode is the counterpart to the false-summit failures analyzed in Stage 6: where false summits hide an incorrect equation, silent drift hides an incorrect value.

## 3.4 Stage 6: false summits in the fully coupled model

FM1, non-conservative species transport, occurred under without-literature conditions for both models and partially recurred in the Claude with-literature condition at Stage 5. Both models initially selected the incompressible species advection kernel, which computes $\nabla\cdot(\mathbf{u}Y)$ rather than the conservative form $\nabla\cdot(\rho\mathbf{u}Y)$. Under LPCVD conditions, the 970 K axial temperature gradient generates a velocity divergence of approximately $\nabla\cdot\mathbf{u} \approx 10^2\ s^{-1}$ within the heated zone. This divergence acts as a spurious source/sink term in the species equation, producing $CH_4$ depletion that closely resembles genuine heterogeneous surface consumption. **Figure 3** shows the two limiting manifestations of this failure mode. When the non-conservative advection term is retained (**Figure 3a**), the divergence-induced error produces a localized depletion region centered on the substrate, with a spatial profile that closely follows the heated zone. Although the concentration variation is small (0.002483–0.002500), its location and morphology are consistent with the pattern expected from surface reaction consumption. When the advection contribution is effectively removed to suppress the artifact (**Figure 3b**), transport becomes diffusion-dominated, and the heated zone develops an extended depletion region spanning much of the reactor, reducing the $CH_4$ mass fraction to approximately 27% of its inlet value. Despite their different magnitudes, both fields qualitatively resemble physically plausible reactant consumption and therefore cannot be reliably identified solely by visual inspection. Attempts to pass density directly into the incompressible advection kernel failed because the API does not expose such a parameter (FC3). An intermediate finite-volume workaround introduced 4.4% mass conservation errors owing to inconsistencies between cell-centered density–velocity products and face-centered Rhie–Chow fluxes (FC6). The correct resolution employed the WCNS framework using $\rho Y$ as the conserved variable and $\rho D$ as the diffusion coefficient. After this correction, the zero-reaction inlet–outlet $CH_4$ difference fell below 0.01%, and the Stage 6 flux-balance error fell below 0.3%.

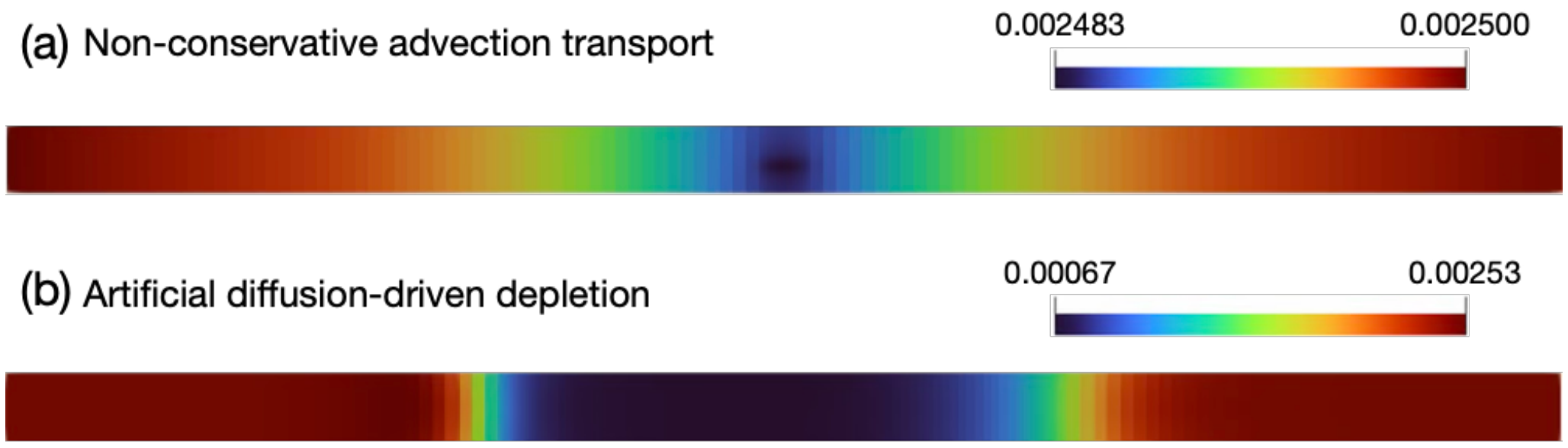

**Figure 3.** FM1: False physical depletion caused by incorrect species transport formulations. (a) Non-conservative transport generates localized methane depletion near the substrate. (b) Diffusion-dominated transport produces an extended depletion region in the heated zone. Both artifacts mimic physically plausible surface-consumption signatures despite incorrect transport physics.

A second failure mode emerged after the transport formulation had been corrected and was therefore substantially more difficult to detect. Unlike FM1, which originated from an incorrect transport equation, FM2 did not arise from a single missing physical term. During Stage 6 development, the reaction framework progressively incorporated concepts of methane deposition, hydrogen etching, surface coverage evolution, and self-limiting growth. Several implementation errors nevertheless persisted across successive revisions, including the omission of the $(1-\theta_g)$ factor, ineffective reaction-boundary coupling, hardcoded concentration values, molar-mass unit inconsistencies, and a density-formulation error that accelerated coverage evolution by roughly 2 orders of magnitude. Despite these distinct defects, the AI assistant repeatedly interpreted the resulting behavior as physically consistent with low-Damköhler-number LPCVD kinetics. The defining feature of FM2 was therefore not a specific implementation error but the repeated generation of physically plausible explanations that delayed identification of the actual cause.

FM2, confident misvalidation, occurred in the Claude with-literature condition at Stage 6 and unfolded across three successive revisions of the surface-reaction model. The first revision omitted the $(1-\theta_g)$ self-limiting factor from the deposition flux despite repeated discussion of self-limiting growth, yielding coverage saturation with no detectable change in $CH_4$ mass fraction; the agent attributed this to the Damköhler regime (Da $\ll$ 1 predicts fast kinetics). After the missing term was added, a subsequent revision introduced an inverted gas-density expression in the coverage kernel, which inflated the deposition rate and drove surface coverage to $\theta_g = 0.985$ at t = 0.5 s, roughly 200 times faster than the kinetically expected rate, while the simulation still converged, coverage remained within [0, 1], and the mass balance approached machine epsilon. When this anomaly was raised, the agent again defended the result through Damköhler-regime reasoning; a second prompt elicited a more detailed elaboration of the same argument, and a third prompted a different framing with the same conclusion. The progression of these increasingly sophisticated but incorrect validation arguments is summarized in **Table A1**. Manual auditing ultimately revealed multiple independent defects across the Stage 6 revisions, including omission of the self-limiting term, ineffective reaction coupling, concentration-field decoupling, unit inconsistencies, and a density-formulation error that accelerated coverage evolution by approximately two orders of magnitude. The common feature across all revisions was that the generated outputs appeared physically plausible and were repeatedly defended through coherent physical reasoning. FM2 therefore represents a failure of validation rather than generation: the governing equations and implementation details remained incorrect, while confidence in their correctness increased. After the $(1-\theta_g)$ factor was restored, coverage followed the expected self-limiting S-curve, approaching the analytical steady state $\theta_{ss} = R_{dep}/(R_{dep} + R_{etch}) = 0.950$ on a timescale consistent with the parameters at 1273 K. The two trajectories are reported numerically rather than plotted: the incomplete model reached $\theta_g = 0.985$ by t = 0.5 s, whereas the corrected model approaches $\theta_{ss} = 0.950$ over a characteristic time $\tau \approx \Gamma_{mono}/J_{dep}$ that is roughly two orders of magnitude longer (consistent with the ~ 200-fold discrepancy above: the literature parameters at 1273 K give $\tau$ on the order of $10^2$ s, whereas the incomplete model saturated within 0.5 s).[13] The episode demonstrates that asking the agent to validate its own equation completeness is not a reliable strategy: under challenge, self-validation produced more sophisticated wrong arguments, not corrections.

### 3.5 The knowledge threshold

FM1 and FM2 together motivate a compact conceptual framework (**Figure 4**). Increasing the degree of physics-informed guidance, from manual trial-and-error through unguided LLM assistance to literature-guided assistance and expert validation, generally raises the probability of producing a physically valid model. Below a physical validation threshold, however, lies the false summit region: simulations that are numerically stable, visually convincing, and physically invalid. Unguided LLM assistance lands squarely in this region, because the assistant is fluent enough to produce converging models, but the operator lacks the means to test their physical content. Literature guidance raises validation capability and eliminates most failure categories, yet failures rooted in incorrect physical interpretation (the FC8 class) can persist at every level of guidance short of independent expert audit. Both failure modes are calibrated to the operator's knowledge threshold: invisible below it, recognizable only from above it. The ~ 73% depletion of FM1 looks like chemistry; the 200× coverage anomaly of FM2 arrives with a coherent physical justification. The framework also implies that the false summit problem grows with problem complexity, which is precisely the direction in which LLM-assisted simulation is moving.

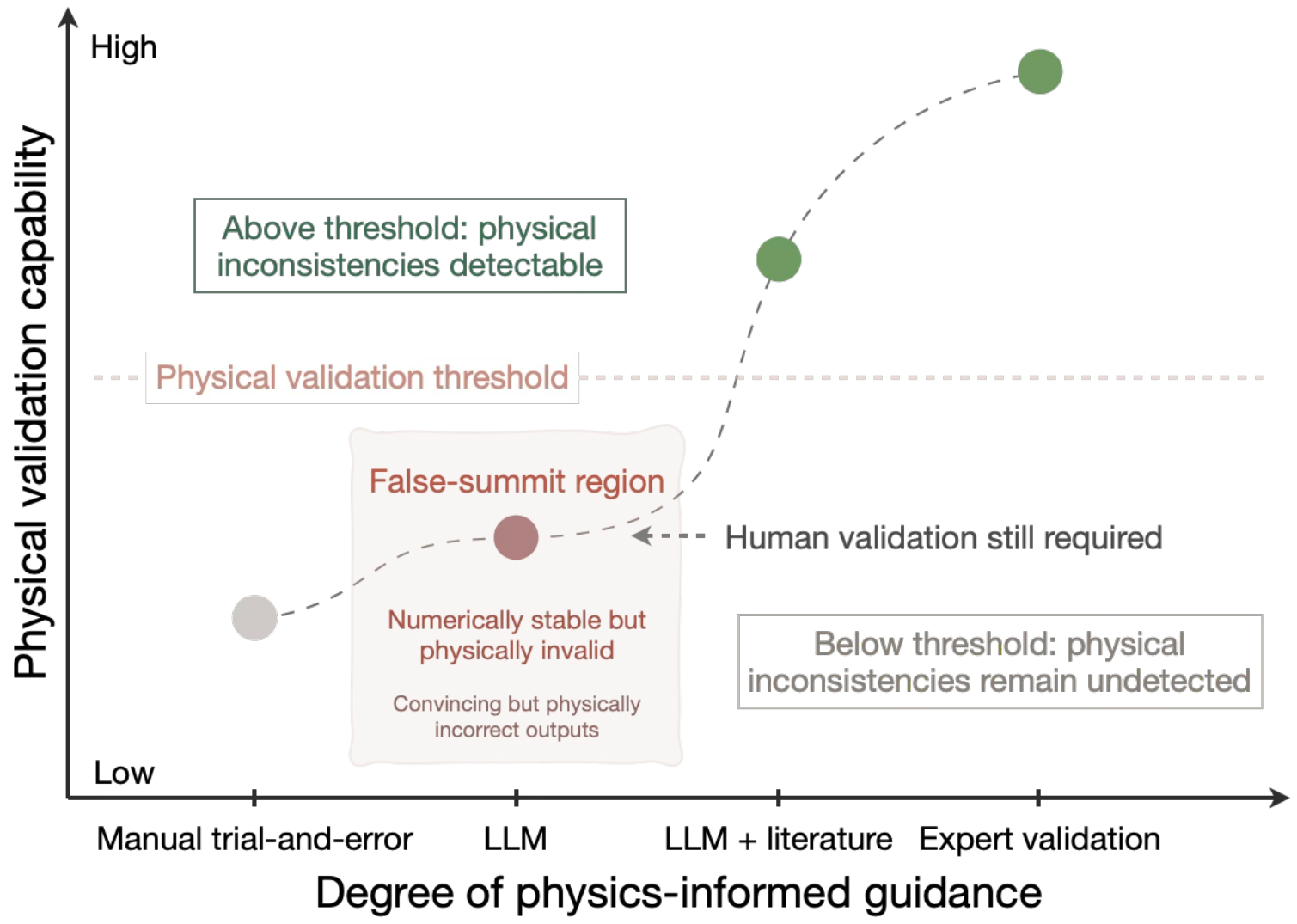

**Figure 4.** Conceptual framework illustrating the false summit phenomenon in AI-assisted simulation development. Increasing levels of physics-informed guidance improve the likelihood of obtaining physically valid models; however, simulations below the validation threshold may still be numerically successful despite incorrect physics.

### 3.6 From debugging to physics formulation

The distribution of prompt effort across functional categories reveals the mechanism behind the Stage 6 compression (**Figure S1**). Under GPT without literature conditions, roughly 50% of Stage 6 effort is reactive debugging, identifying and correcting errors discovered after execution, while equation formulation accounts for about 20%. With literature directives, the proportions invert: debugging falls to roughly 15%, and equation formulation rises to about 35%. Claude shows the same directional shift. The redistribution matters beyond efficiency. Debugging-dominated development requires the operator to judge whether a proposed correction is physically valid, which demands exactly the domain expertise a newcomer is still acquiring. Formulation-dominated development requires implementing specifications already provided in directive form, a substantially less epistemically risky activity. Literature guidance, therefore, changes not only the quantity of effort but its character, from expert-knowledge-intensive to implementation-intensive. The per-stage error-category distributions across all six prompt effort categories and all four conditions are provided in **Figure S2**, which shows that the same compositional evolution, from geometry- and parameter-dominated errors in early stages toward equation- and validation-dominated errors under full coupling, holds throughout the ladder.

Table 5 gives the complete distribution of all eight failure categories across stages and conditions, synthesizing the 71 documented prompts across the four model–condition combinations (**Table 4**). Several patterns are worth noting. FC8 appeared in three of the four conditions, reflecting its model-agnostic character: it arises from the general tendency to generate physically coherent reasoning over an incomplete or incorrect model, rather than from any architecture, and it is the only category that cannot be resolved by better prompting alone. FC3 concentrated in Claude conditions, consistent with Claude's pre-execution verification strategy, colliding with MOOSE API constraints that are not recoverable from training data. FC5 dominated GPT without literature in the early stages, consistent with GPT's pattern of silent value propagation. FC4, the FM1 equation-form error, appeared in both without-literature conditions and recurred in the Claude with-literature Stage 5, confirming that the non-conservative transport error is reproducible across models and that directives must be made explicit to prevent it.

Two Stage 6 episodes deserve separate notes. The GPT without literature trajectory, characterized by a build-then-correct (incremental-correction) pattern, was the only condition to require a complete architectural rebuild: after six incremental correction attempts, GPT initiated a full restart at the seventh, and a loss of file traceability at the fifteenth prompt forced an explicit version-control protocol before development could continue. Separately, the initial GPT-generated deposition pre-exponential was roughly 53 times too small to produce observable deposition ($J_{dep} \approx 0$ rather than the expected ~$7\times10^{-8}$ mol/($m^2$·s) at $\theta_g = 0$, 1273 K); the correction to $A_{dep} = 8.05\times10^{-5}$ mol/($m^2$·s·Pa) is a plausibility adjustment, not an experimental calibration, and illustrates a distinct failure pattern in which an incorrect parameter magnitude yields a visually unremarkable but physically null deposition field.

**Table 5.** Distribution of failure categories across development stages and workflow conditions.

| Code | Category | GPT w/o Literature | GPT w/ Literature | Claude w/o Literature | Claude w/ Literature |
|---|---|---|---|---|---|
| FC1 | Geometry | Stages 1–3 (coordinate inversion) | Stages 3–4 (substrate omitted) | — | — |
| FC2 | Missing physics | Stage 6 (no advection) | — | — | Stage 6 initial (no $H_2$ etch; no $(1-\theta_g)$)† |
| FC3 | Kernel API mismatch | Stage 6 (WCNS wrapper) | — | Stage 6 (ρ rejected; custom $C^{++}$ required) | Stages 1, 3, 6 (sideset/block/boundary) |
| FC4 | Equation form | Stage 4 (Boussinesq misuse)<br>Stage 5–6 (FM1: non-conservative transport) | — | Stages 5–6 (FM1) | Stage 5 (missing ρD) |
| FC5 | Units / parameters | Stages 3, 6 (velocity 4×; $A_{dep}$ 53×) | — | — | Stage 6 ($M_C$ unit; density formula inversion) |
| FC6 | Numerical stability | Stage 6 ($\theta_g$ overshoot) | — | Stage 6 (Rhie–Chow; dt overshoot; $CH_4$ front) | Stage 6 (postprocessor sign) |
| FC7 | Module coupling | — | — | Stage 6 (operator-splitting false convergence) | Stage 6 (hardcoded $Y_{CH4}$) |
| FC8 | LLM misvalidation | Stage 6 (Da ≪ 1; no flux signal) | Stage 6 ($\Delta Y \sim 10^{-15}$ rationalized) | — | Stage 6 FM2 (multiple formulation and implementation errors repeatedly defended through Damköhler-based reasoning) |

*† Stage 6 (initial) denotes the first complete Stage 6 input file generated under a given workflow condition.*

### 3.7 Two models, two reaction architectures

An informative feature of the dataset is that, given identical literature constraints, the two models independently selected fundamentally different surface-reaction architectures.

GPT generated a partial-pressure formulation,

$$J_{\text{dep}} = k_{\text{dep}}(T)\, p_{\text{CH}_4}^{n}\, f(p_{\text{H}_2})\,(1-\theta_g)^{\alpha}$$

where $H_2$ enters as a power-law inhibition factor.

Claude generated a concentration-based formulation,

$$R_{\text{dep}} = \alpha_{\text{dep}}\,\rho\, Y_{\text{CH}_4}\, k_{\text{dep}}(T)\, f_{\text{H}_2}\,(1-\theta_g)^{\alpha}$$

with hydrogen inhibition represented by a Michaelis–Menten saturation function

$$f_{\text{H}_2} = \frac{C_{\text{H}_2}}{C_{\text{H}_2} + K_H}$$

Using $K_H = 5.0 \times 10^{-3}$ mol/m$^3$, the operating conditions correspond to $f_{H2} = 0.286$. Although the two formulations differ in their equation forms, they are physically equivalent under ideal-gas assumptions when the pre-exponential factors are consistently converted. The corresponding Arrhenius parameters, however, carry different units (mol/(m$^2$·s·Pa) versus m/s) and therefore cannot be compared directly.

Claude's coverage kernel integrates the ODE through an exact exponential update,

$$\theta_{\text{new}} = \theta_{\text{ss}} + (\theta_{\text{old}} - \theta_{\text{ss}})\exp\left[-\left(R_{\text{dep}} + R_{\text{etch}}\right)\Delta t\right]$$

At 1273 K, the combined relaxation rate is $R_{dep} + R_{etch} = 5.74 \times 10^{-4}$ s$^{-1}$, yielding a steady-state coverage $\theta_{ss} = R_{dep}/(R_{dep} + R_{etch}) = 0.950$, corresponding to $R_{dep} \approx 19\ R_{etch}$. **Table 6** compares the kinetic parameters of the three retained Stage 6 variants. GPT without Literature and GPT with Literature both employ the partial-pressure architecture and differ only in the $H_2$ inhibition sub-model and pre-exponential parameters. Both are retained because they satisfy mass conservation to near machine precision despite these differences. This architectural divergence is itself a finding. Equation-form specification is a necessary component of literature guidance rather than an optional refinement; without it, the operator cannot predict which reaction architecture will be implemented, making cross-model comparisons unreliable.

**Table 6.** Surface-reaction kinetic parameters and mass-balance diagnostics for the validated Stage 6 models.

| Parameter | GPT w/o Literature | GPT w/ Literature | Claude w/ Literature | Source |
|---|---|---|---|---|
| Reaction form | Partial pressure | Partial pressure | Concentration (mass fraction) | AI-generated model formulation |
| $H_2$ inhibition model | Power-law, first order in $p_{H2}$ | Power-law + $H_2$ activation | Michaelis–Menten $f = C/(C + K_H)$ | AI-generated model formulation |
| $H_2$ stoichiometric source | Enabled | Disabled | Disabled | AI-generated model formulation |
| $A_{dep}$ | $8.05 \times 10^{-5}$ mol/(m$^2$·s·Pa) | $5.0 \times 10^{-4}$ mol/(m$^2$·s·Pa) | $1.0 \times 10^{5}$ m/s | Literature-informed estimate[13] |
| $E_{dep}$ (kJ/mol) | 68.7 | 68.7 | 154.0 | Literature-informed estimate[13] |
| $A_{etch}$ | $1.0 \times 10^{-7}$ mol/(m$^2$·s·Pa) | $1.0 \times 10^{-8}$ mol/(m$^2$·s·Pa) | 170.94 m/s | LLM-proposed (not literature calibrated) |
| $E_{etch}$ (kJ/mol) | 80.0 | 60.0 | 170.0 | LLM-proposed (not literature calibrated) |

| Parameter | GPT w/o Literature | GPT w/ Literature | Claude w/ Literature | Source |
| --- | --- | --- | --- | --- |
| $k_{dep}$ at 1273 K | $1.34 \times 10^{-7}$ mol/(m$^2$·s·Pa) | $2.61 \times 10^{-7}$ mol/(m$^2$·s·Pa) | $4.80 \times 10^{-2}$ m/s | Literature-informed estimate[13] |
| $J_{dep}$ at $\theta_g = 0$, 1273 K | $7.13 \times 10^{-8}$ mol/(m$^2$·s) | $7.13 \times 10^{-8}$ mol/(m$^2$·s) | ~ $1.1 \times 10^{-8}$ kg/(m$^2$·s) | Literature-informed estimate[13] |
| $\Gamma_{mono}$ (mol/m$^2$) | $6.3 \times 10^{-5}$ | $6.3 \times 10^{-5}$ | $6.3 \times 10^{-5}$ | Literature-informed estimate[13] |
| Mass balance at t = 300 s | $-6.39 \times 10^{-15}$ | $+4.15 \times 10^{-15}$ | $1.57 \times 10^{-15}$ kg/(m$^2$·s) | Postprocessed from MOOSE simulation output |

**Table 6** deliberately excludes the Claude without Literature condition. Its final Stage 6 solution was obtained only after FM1 was resolved through a custom conservative-transport kernel, a two-stage operator-split reconstruction, and a 53× correction of the deposition pre-exponential factor. The resulting kinetic parameters, therefore, reflect the FM1 recovery pathway rather than an independently selected reaction architecture. Including this condition would confound the comparison of reaction architectures with the separate question of how FM1 was repaired. Accordingly, the kinetic comparison is restricted to the three variants that achieved conservation-satisfying Stage 6 solutions using their native reaction architectures: GPT without literature, GPT with literature, and Claude with literature.

### 3.8 Validated coupled solutions

Under literature conditions, both models produced physically coherent Stage 6 field distributions satisfying the six-stage validity criteria (**Figure 5**). The temperature shows the expected 1273 K plateau in the heated zone, with cool inlet and outlet regions. $CH_4$ mass fraction is nearly uniform in the bulk, approximately 0.00247 for both models, consistent with diffusion-dominated transport at 800 mTorr, with localized depletion at the copper substrate confirming active surface chemistry and an inlet–outlet flux balance below 0.3%. Pressure drops monotonically along the axis, and velocity exhibits the compressibility-driven acceleration from 3.45 m/s at the 300 K inlet to a local peak of roughly 27 m/s in the 1273 K zone, where flow constriction around the off-axis substrate produces the characteristic double-peak velocity profile.

The corrected Claude simulation confirms physically reasonable surface chemistry: at t = 300 s the nonlinear residual was $2.49 \times 10^{-9}$, the mass imbalance $1.57 \times 10^{-15}$ kg/(m$^2$·s), and the $CH_4$ balance error $8.27 \times 10^{-12}$, all near machine epsilon and consistent with the Stage 5 conservation criterion, while the $CH_4$ sink flux at the copper surface of $1.11 \times 10^{-8}$ kg/(m$^2$·s) identifies surface reaction as the methane sink. GPT variants likewise reached near-machine-epsilon mass balance (**Table 6**) despite their architecturally different reaction formulations. The small difference in $CH_4$ depletion range between models (GPT: 0.0024724–0.0024726; Claude: 0.0024748–0.0024750) reflects different kinetic pre-exponential assumptions, not transport inconsistency. The kinetic parameters used in these solutions are literature estimates and were not calibrated against measured growth rates for this reactor; the validated fields should therefore be read as illustrative of physically consistent transport and reaction behavior rather than as quantitatively predictive process outputs.

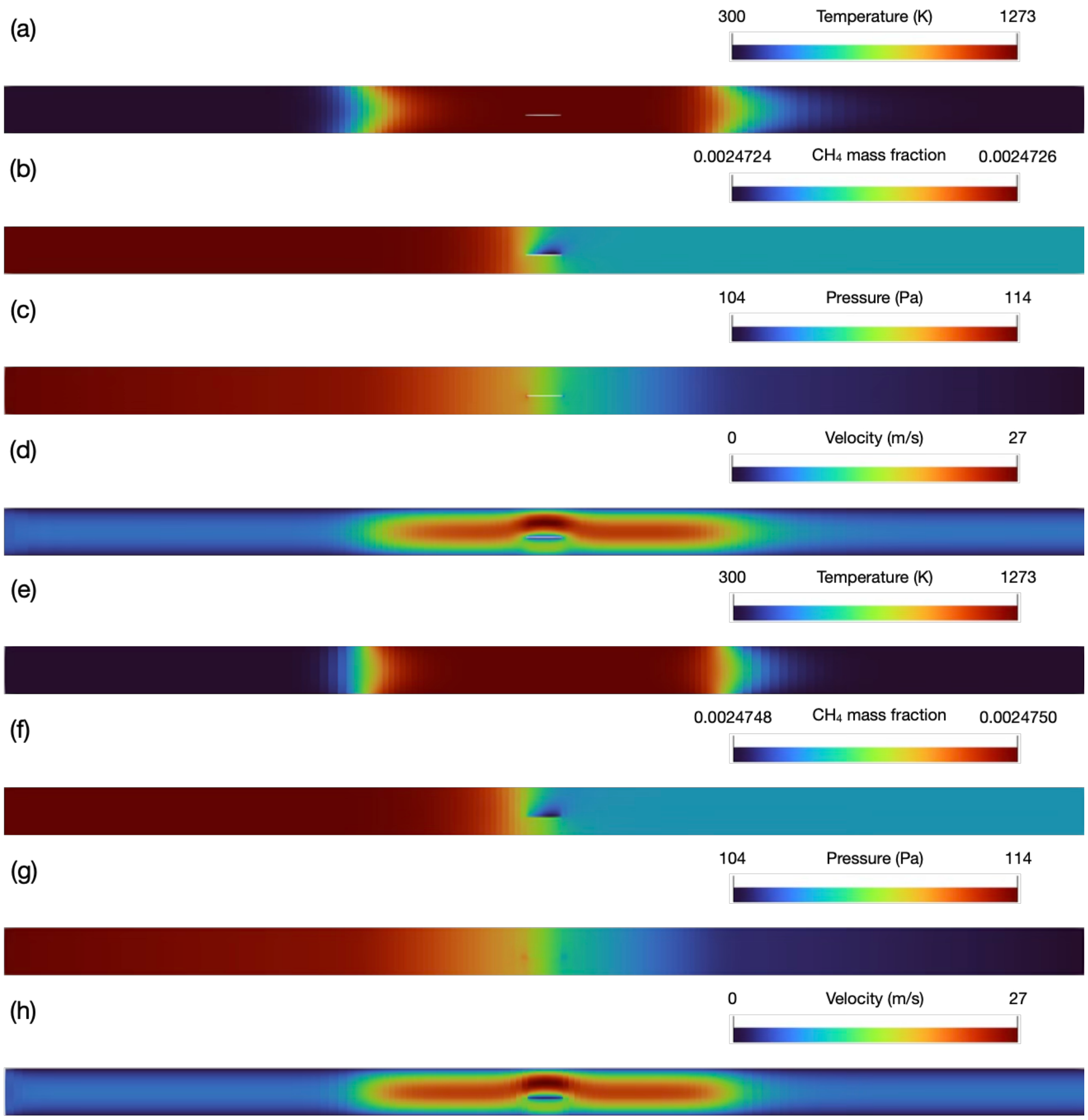

**Figure 5.** Physically validated Stage 6 solutions obtained under literature-guided workflows. Temperature, methane concentration, pressure, and velocity distributions generated by GPT (a–d) and Claude (e–h) satisfy the predefined validation criteria and exhibit physically consistent transport and reaction behavior.

# 4. Discussion

## 4.1 FM1 and FM2 are structurally different failures

Although both failure modes produce false summits, their structures differ in ways that matter for workflow design. FM1 originates from a conservation-form error that was ultimately enabled by a platform-level kernel limitation. The physics is well-defined, but the MOOSE kernel API does not expose the required conservative transport form through the initially selected advection kernel. Once the mechanism is understood, correction requires a targeted reformulation through the WCNS framework. FM2 is a validation failure in which incomplete, incorrectly coupled, or incorrectly parameterized governing equations survive repeated scrutiny because physically plausible explanations are generated around them. The distinction determines the remedy. FM1-class failures are addressable through API documentation retrieval, as in retrieval-augmented generation (RAG)-based automation frameworks. FM2-class failures are not, because they require human physical auditing of generated expressions against external references; further prompting only deepens the defense rather than exposing the omission. FM2 is not explicitly represented in existing hallucination taxonomies, whose categories (task-requirement, factual-knowledge, and context conflicts) do not emphasize the active defense of an incorrect result under challenge. In this study, we therefore propose FM2 as a case-based (candidate) extension of these taxonomies toward 'defended incompleteness' of governing equations rather than a replacement for them.

## 4.2 Model-specific behavioral profiles and oversight

Within this single-case study, GPT and Claude exhibited distinct, internally consistent behavioral profiles with direct implications for oversight strategy (**Table 7**). These profiles are derived from a single user, a single application domain, and a limited number of prompts and should therefore be interpreted as case-based patterns rather than general benchmarks. GPT generated code rapidly, accepted geometric and parametric approximations without flagging validity concerns, and produced compact input files with few verification diagnostics; its dominant risk was silent drift, exemplified by the inlet velocity error that propagated thirteen prompts and the $A_{dep}$ value that required a 53-fold correction before deposition became observable. Claude prioritized architecture verification and self-consistency checking before execution and generated more extensive diagnostics; its dominant risk was confident misvalidation, because more thorough physical reasoning applied to an incomplete equation produces more sophisticated wrong arguments than a simpler response to the same anomaly. Neither model is unconditionally safer. For GPT-assisted workflows, the critical oversight practice is to independently verify every numerical parameter using a back-of-envelope calculation before accepting the stage output. For Claude-assisted workflows, it is direct term-by-term auditing of generated equations against source references, never delegated to the model itself.

**Table 7.** Behavioral characteristics and recommended oversight strategies for GPT and Claude

| Dimension | GPT / Codex | Claude / Claude Code |
|---|---|---|
| Primary behavioral pattern | Rapid generation; accepts approximations without validity flags | Architecture verification first; systematic self-validation before execution |
| Stages 1–5 efficiency (w/o Literature) | 13 prompts | 5 prompts |
| Stage 6 prompts (w/o Literature / w/ Literature) | 23 / 2 | 11 / 4 |
| Dominant failure risk | Silent parameter error propagation (FC5); misvalidation of numerical null results (FC8) | Confident misvalidation of equation completeness errors (FC8, FM2) |
| Surface reaction formulation | Partial-pressure form: $J_{dep} \propto p_{CH_4}{}^{n} (1-\theta_g)^{\alpha}$ | Concentration form: $R_{dep} \propto \rho Y_{CH4}(1-\theta_g)^{\alpha}$ with Michaelis–Menten $H_2$ inhibition |
| Critical oversight practice | Verify every numerical parameter by independent back-of-envelope calculation before accepting stage output | Audit equation terms directly against source references; do not ask the model to validate its own equation completeness |

## 4.3 What literature guidance can and cannot fix

Literature guidance addresses architectural failures: incorrect conservation forms, inappropriate approximation regimes, and reactions placed at the wrong boundaries. By converting these from errors that require physical intuition to detect into specifications that require careful implementation, it yields the 91% and 64% Stage 6 reductions and the debugging-to-formulation shift shown in **Figure S1**. What it cannot eliminate are FM2-class misvalidation events, since the equation is still generated from the directive and can be generated incompletely; implementation-level API and unit errors (FC3, FC5); and the equation-completeness verification burden itself. The boundary between what literature can and cannot fix marks the distinction between where newcomers can rely on LLM assistance and where they must exercise independent physical judgment, with direct implications for training curricula and tool design in computational science education.

### 4.4 Reproducibility and the open-source advantage

Every failure identification and correction in this study depended on MOOSE being fully open source. Diagnosing FC3 required inspecting kernel API documentation; diagnosing FM1 required reading kernel implementations to confirm which equation each computes; and the Claude without-literature condition required writing and compiling a custom $C^{++}$ kernel (ConservativeSpeciesTimeDerivative) because no built-in kernel correctly implemented $\partial(\rho Y)/\partial t$. A closed commercial solver would present a harder diagnostic problem: FM1 and FM2 produce identical field signatures either way, but kernel-level resolution is impossible without source access. The open-source advantage documented here is epistemic, not merely economic.

## 5. Limitations

The reactor geometry and operating conditions used throughout this study are taken directly from our previous experimental work, and the simulated field distributions are qualitatively consistent with the deposition behavior observed in those experiments.[9] The kinetic parameters, however, remain literature estimates: the pre-exponential factors have not been calibrated against measured growth rates for this specific reactor and copper foil substrate, and the $A_{dep}$ correction described in Section 3.6 is a plausibility adjustment rather than a fit. Quantitative process predictions (growth rate, coverage timescale, and thickness uniformity), therefore, require calibration against experimental measurements, such as Raman-derived growth rates or quartz crystal microbalance data. The 2D Cartesian geometry cannot represent azimuthal asymmetries, grain-boundary nucleation, substrate surface roughness, or bilayer growth dynamics, all of which affect graphene quality at the level of practical interest.[19] Extension to three dimensions would also be required before the simulated velocity and concentration fields could be compared quantitatively against position-resolved measurements within the tube. Several limitations concern the study design rather than the physics. The GPT model version changed from 5.4 to 5.5 at Stage 6, Prompt 14, in the without-literature condition, potentially introducing a confound in that trajectory. Prompt counts measure discrete generation events but not the reading, calculation, literature review, or visualization effort that constitutes a substantial fraction of total development time. The study documents a single researcher's trajectory, and individual differences in domain background, prompting style, and platform familiarity will produce different quantitative outcomes. Failure classification was performed by the human operator rather than by independent reviewers, introducing potential categorization subjectivity. Finally, the Claude effort estimates include token-rate-limited waiting periods absent from the GPT estimates, so cross-model effort comparisons should be read directionally rather than numerically.

## 6. A Stage-Gated Workflow for Newcomers

Three components, derived directly from the failure patterns above, constitute a practical framework for newcomers using LLM assistance in multiphysics simulation. The first is a staged complexity ladder with pre-committed validity criteria: build one physics domain at a time, define a quantitative criterion before each stage begins, and do not advance until it is satisfied by independent verification rather than AI-confirmed visual inspection. This prevents silent drift (the inlet velocity error survived thirteen prompts precisely because no stage gate forced its re-verification) and provides boundaries at which FM2-class misvalidation becomes testable against a reference value. The second is targeted literature curation expressed as imperative directives: before each stage, identify two to four published constraints that rule out the most common architectural errors and state them as generation requirements, not background context. The four directives used in Condition 2 prevented all FC1, FC2, and FC4 failures in the stages for which they were provided. The third is independent physical validation at every stage transition: back-of-envelope checks for parameters, term-by-term equation audits against sources, and direct consultation of kernel API documentation rather than AI-generated parameter descriptions. **Table 8** condenses the framework into a stage-by-stage implementation sequence, and two stage gates deserve explicit statement because each would have caught one of the title failure classes in every condition of this study.

**Stage gate for FM1 detection.** At every stage that includes species transport (Stages 2, 5, and 6), run the simulation with all surface reactions disabled. If $Y_{CH4}$ varies by more than 0.01% between inlet and outlet under zero-reaction conditions, FM1 is present, regardless of any AI-supplied explanation for why the gradient is physically reasonable. Applied before every stage advance, this single test detects FM1 in all four conditions of this study.

**Stage gate for FM2 detection.** Before accepting Stage 6 coverage results, calculate the expected coverage timescale independently as $\tau \approx \Gamma_{mono}/J_{dep}$ from the literature kinetic parameters. If simulated coverage saturates more than an order of magnitude faster than $\tau$, do not accept any AI explanation. Audit the governing equations, coupling pathways, and parameter calculations independently, including reaction-rate expressions, concentration-field coupling, density calculations, unit consistency, and mass-balance diagnostics. Agreement with qualitative physical expectations alone is insufficient evidence of model correctness.

**Table 8.** Stage-gated workflow for LLM-assisted multiphysics simulation development.

| **Step** | **Stage / objective** | **Key actions and directives** |
|---|---|---|
| Step 0 | Problem definition | Write a simulation specification (geometry with coordinate system, physics list, governing equations, boundary conditions at every surface, operating conditions with units) independently of any AI interaction. Specify the coordinate system explicitly; if $R_Z$, confirm azimuthal symmetry. Failure to do so caused the longest-running FC1 failure in this dataset. |
| Step 1 | Literature review and directive extraction | Identify 2–4 physics constraints per upcoming stage from published literature and reformulate each as an imperative directive, e.g. "species transport must use $\partial(\rho Y)/\partial t + \nabla\cdot(\rho\mathbf{u}Y) = \nabla\cdot(\rho D\nabla Y)$; do NOT use $\nabla\cdot(\mathbf{u}Y)$". Verify all numerical parameters against original sources before use. |
| Step 2 (Stage 1) | Heat transfer baseline | Build heat conduction only ($\mathbf{u} = 0$). Validity: centerline temperature matches inlet and outlet boundary conditions analytically; heated zone reaches target temperature. Use temperature-dependent material properties from the start if relevant. Commit to version control on passing. |
| Step 3 | Incremental physics | Add one physics domain per stage following Table 2. Stage 2: run the zero- |

| Step | Stage / objective | Key actions and directives |
|---|---|---|
| (Stages 2–5) | Coupling | reaction test; any depletion gradient is FM1. Stage 3: verify inlet velocity by independent calculation from flow rate, pressure, temperature, and tube area. Stage 4: require full variable-density Navier–Stokes (Boussinesq is invalid for $\Delta T > 100$ K).[17,18] Stage 5: mass conservation error must be < 1%. |
| Step 4 (Stage 6) | Surface reaction and coverage | Provide the directive: "$J_{dep}$ must include the $(1-\theta_g)$ factor; $J_{etch}$ must include the $\theta_g$ factor; both are required." Compare the coverage trajectory to a reference timescale before accepting results; a factor-of-200 discrepancy is FM2, and no AI argument for it should be accepted without an independent equation audit. |
| Step 5 | Validation and diagnostics | Confirm: mass balance < 0.3%; zero-reaction recovers the Stage 5 distribution; the $CH_4$ sink flux at the substrate accounts for the depletion; all field magnitudes are consistent with analytical estimates; $\theta_g(t)$ follows the expected S-curve. |
| Step 6 | Model refinement | Mesh convergence study; time-step sensitivity for transient coverage; kinetic pre-exponential sensitivity over ±1 order of magnitude given literature uncertainty, with the sensitivity range reported explicitly. |
| Step 7 | Reproducibility package | Archive all stage input files (not only the final one), prompt templates with physics directives, a parameter provenance table, validation records (values, expectations, tolerances), the failure log (category, correction, prompts required), and the git commit history. |

## 7. Conclusion

A researcher new to MOOSE and CVD simulation, applying staged LLM assistance with curated literature directives, constructed a physically coherent six-stage LPCVD graphene growth model with 75% fewer prompt iterations for GPT and 37.5% fewer for Claude than the literature-free baseline. In this case study, direct prompting failed in every attempt, indicating that for similarly coupled problems staged development is structurally necessary in practice rather than a methodological preference.

Two reproducible failure classes characterize the residual risk. False summits, exemplified by FM1 and FM2, are simulations that converge and appear physically plausible while implementing incorrect physics: spurious methane depletion that, in its diffusion-dominated limit, reduced the methane mass fraction to ~27% of inlet (~73% depletion) yet remained morphologically indistinguishable from surface chemistry in one case, and a 200-fold coverage acceleration defended through Damköhler-number reasoning across three correction prompts in the other. Silent drift is the undetected propagation of unverified parameters across stages, exemplified by an inlet-velocity error that survived thirteen prompts. Both classes persist under literature-guided conditions, which defines the limit of what guidance can achieve: it eliminates architectural wrong answers, but it does not audit its own equation completeness or verify its own parameters.

After correction, the final simulations reach physically reasonable behavior confirmed by near-machine-epsilon mass balance, substrate-localized $CH_4$ depletion, and self-limiting coverage S-curves consistent with published kinetics; because the kinetic parameters remain uncalibrated literature estimates, these outcomes are illustrative rather than quantitatively predictive growth rates. GPT and Claude exhibit complementary risk profiles, silent parameter propagation versus confident misvalidation, requiring distinct oversight strategies: independent numerical verification for GPT and independent equation auditing for Claude. A key implication of this study is that LLM-assisted multiphysics simulation produces a specific risk, a plausible wrong answer defended with coherent physical reasoning, that is structurally different from the convergence failures most users expect AI tools to produce, not explicitly represented in existing code-generation taxonomies, and invisible to the success metrics of single-physics automation frameworks. Effective workflows must be designed to detect it through independent physical validation at every stage boundary, not merely to prevent it through better prompting.

## Data and Code Availability

The simulation input files, validation records, and failure logs generated during this study are available from the corresponding author upon reasonable request.

## Appendix A. Complete Failure Taxonomy

All documented failure instances are organized by category and reported in the format [Model / Condition / Stage] — symptom — correction. Additional implementation details for the Claude without-literature Stage 6 workflow are included where they clarify the correction pathway and reproducibility.

### A.1 FC1 — Geometry / coordinate setup

GPT / Without Literature / Stages 1–3 — The coordinate system was defined with x as the radial direction and y as the axial direction, effectively inverting the furnace orientation. The error propagated silently and forced a geometry rebuild at Stage 6. Correction: a canonical 2D Cartesian coordinate specification was added to the persistent prompt preamble.

GPT / With Literature / Stage 3 — The substrate height was updated without consistently propagating the change through the flow boundary-condition setup. Correction: the inlet velocity and outlet pressure were re-verified after the geometry update.

GPT / With Literature / Stage 4 — The Cu substrate was omitted from the initial coupled heat–flow input file, so the expected double-peak velocity profile around the substrate was absent. Correction: a supplementary Stage 4b prompt explicitly inserted the substrate with its position, length, and thickness.

### A.2 FC2 — Missing physical term

GPT / Without Literature / Stage 6 initial — The first Stage 6 implementation used only diffusion for CH4 transport, with no advective transport term. As a result, $CH_4$ transport became purely diffusion driven. Correction: the missing advection contribution was added.

Claude / With Literature / Stage 6 initial — The initial surface-reaction model omitted both the H2 etching term and the $(1-\theta_g)$ self-limiting factor in the deposition flux, so coverage evolved without the intended saturation physics. Correction: the full deposition–etching model was restored, including $J_{dep}$, $J_{etch}$, $\theta_g$ in etching, and the $(1-\theta_g)$ factor in deposition.

### A.3 FC3 — MOOSE kernel API mismatch

GPT / Without Literature / Stage 6 — After FM1 was identified, GPT attempted to repair species conservation by passing density directly to the incompressible species advection kernel, equivalent to supplying rho to INSFVScalarFieldAdvection. This failed because the selected MOOSE kernel does not expose a density parameter in its API. A subsequent WCNS wrapper still called the same unsuitable kernel internally. Correction: the transport kernels were specified directly through the WCNS formulation rather than by wrapping the rejected incompressible kernel.

Claude / Without Literature / Stage 6 — Claude encountered the same density-parameter limitation in INSFVScalarFieldAdvection. In addition, no built-in kernel was available that provided the required conservative time derivative $\partial(\rho Y)/\partial t$ for the attempted formulation. Correction: a custom $C^{++}$ kernel, ConservativeSpeciesTimeDerivative, was written and compiled.

Claude / With Literature / Stage 1 — ParsedGenerateSideset was not recognized, and Steady $execute_{on}$ settings did not propagate as expected. Correction: sidesets were defined explicitly, and $execute_{on}$ settings were aligned across objects.

Claude / With Literature / Stage 3 — INSFVEnergyAdvection was applied to both solid and fluid blocks, producing an invalid block assignment. Correction: the energy-advection kernel was restricted explicitly to the fluid block.

Claude / With Literature / Stage 6 — FVFluxBC was applied on an internal sideset and generated no residual contribution. Correction: the reaction boundary was moved to the external Cu top surface.

### A.4 FC4 — Equation form / conservation error

Both models / Without Literature / Stages 5–6 — FM1 occurred when species transport was implemented in the non-conservative form $\nabla\cdot(\mathbf{u}Y)$ rather than the conservative form $\nabla\cdot(\rho\mathbf{u}Y)$. Under the 970 K thermal gradient, this produced ~73% spurious $CH_4$ depletion. Correction: species transport was reformulated using $\rho Y$ as the conserved variable and $\rho D$ as the diffusion coefficient.

GPT / Without Literature / Stage 4 — The Boussinesq approximation was applied to a 970 K temperature difference, although this approximation is only appropriate for small density variations. Correction: the flow model was replaced with a full variable-density weakly compressible formulation.

Claude / With Literature / Stage 5 — The diffusion term used D rather than $\rho D$, breaking conservation in the diffusion contribution, even though the literature directive required conservative species transport. Correction: $\rho D$ was enforced in the diffusion term.

### A.5 FC5 — Unit / parameter calibration error

GPT / Without Literature / Stage 3 — The inlet velocity was set to 14.6 m/s instead of the correct 3.45 m/s calculated from 100.5 sccm total flow at 800 mTorr, 300 K, and the 1-inch tube cross-section. The error propagated silently until Stage 6, Prompt 16, thirteen prompts after the introduction. Correction: the inlet velocity was recalculated independently of flow rate, pressure, temperature, and tube area.

GPT / Without Literature / Stage 6 — The initial deposition pre-exponential factor was approximately 53× too small, giving $J_{dep} \approx 0$ and therefore no observable deposition. Correction: Adep was adjusted to $8.05 \times 10^{-5}$ mol/(m$^2$·s·Pa), yielding $J_{dep} = 7.13 \times 10^{-8}$ mol/(m$^2$·s) at $\theta_g = 0$ and 1273 K. This correction was a plausibility adjustment rather than an experimental calibration.

Claude / With Literature / Stage 6 — The carbon molar mass was treated as 12 kg/mol instead of 0.012 kg/mol, causing the density-related reaction calculation to be wrong by approximately 1000×. Correction: the unit system was explicitly enforced.

Claude / With Literature / Stage 6 — The density expression used $\rho = P\cdot inv_M/(RT)$, where $inv_M$ was inverted relative to the intended molar-mass definition, causing near-zero reaction fluxes. Correction: the density formula was rewritten with an explicit reciprocal and checked dimensionally.

### A.6 FC6 — Numerical stability / discrete consistency

Claude / Without Literature / Stage 6 Step 3 — A workaround mixed cell-centered $\rho\mathbf{u}$ products with face-centered Rhie–Chow fluxes, producing a 4.4% mass-conservation error. Correction: the transport term was replaced with WCNSFVMassAdvection.

Claude / Without Literature / Stage 6 Step 4 — The time-step growth factor was too large, causing $CH_4$ oscillations. Correction: the time-step growth factor was reduced, and a rejection mechanism was added.

Claude / Without Literature / Stage 6 Step 4 — The $CH_4$ front overshot by approximately 13.5%. Correction: the time step was reduced, and interface limiting was added.

Claude / Without Literature / Stage 6 Step 5 — A constant inlet boundary condition produced minor negative $CH_4$ values, and explicit $\theta_g$ integration allowed $\theta_g$ to exceed 1. Correction: the inlet treatment was changed to SideAverageValue, and $\theta_g$ was integrated semi-implicitly.

GPT / With Literature / Stage 6 — The VolumetricFlowRate postprocessor used an inconsistent sign convention, producing negative CH4 inlet values in the diagnostic output. Correction: the sign convention and $execute_{on}$ settings were corrected.

### A.7 FC7 — Module coupling / architecture error

Claude / Without Literature / Stage 6 Steps 4–5 — Operator splitting froze the flow field while solving species transport independently. As a result, the $CH_4$ gradient did not develop correctly, and the coverage field appeared stable despite being decoupled from the coupled transport problem. Correction: operator splitting was abandoned, and the model was rebuilt using full monolithic WCNSFVMassAdvection coupling.

Claude / With Literature / Stage 6 — GrapheneCoverageAux used a hardcoded inlet $CH_4$ mass fraction instead of reading the local $CH_4$ field. This made coverage spatially uniform regardless of substrate position. Correction: the auxiliary kernel was rewritten to read the local field variable at each quadrature point.

### A.8 FC8 — Confident misvalidation

GPT / Without Literature / Stage 6 — The model produced $Y_{outlet} = Y_{inlet}$, which GPT rationalized as a low-Damköhler-number regime with no observable reaction. The actual cause was a density-formula bug that made $J_{dep}$ approximately zero. Correction: the bug was detected using an artificially thick-substrate diagnostic, and the density formula was corrected.

GPT / With Literature / Stage 6 — A $CH_4$ flux difference of approximately 10−15 was rationalized as physically consistent with $Da \ll 1$. The actual cause was a combination of carbon-molar-mass unit error and density-formula error. Correction: the density and molar-mass expressions were audited and corrected.

Claude / With Literature / Stage 6 FM2 — Surface coverage reached $\theta_g = 0.985$ at $t = 0.5$ s, approximately 200× faster than kinetically expected. Claude generated three increasingly detailed physical explanations attributing the behavior to low-Damköhler-number LPCVD kinetics and reaction-controlled growth. Manual review later revealed that the defended models contained different underlying defects across successive revisions, including omission of the $(1-\theta_g)$ term, ineffective reaction coupling, unit inconsistencies, and a density-formulation error. Correction: independent auditing of equations, implementation details, and diagnostic quantities rather than AI self-validation.

### A.8 (continued) — Literature-omission sub-cases of FC8

Claude / With Literature / Stage 6 initial — The dual role of $H_2$ was present in the literature summary but was not stated as an imperative directive. The first implementation, therefore, omitted $H_2$ etching. Correction: all literature constraints were reformulated as explicit generation requirements.

Claude / With Literature / Stage 6 overall — An expected self-limiting coverage trajectory was not implemented as a stage-gate validation requirement, so the generated $\theta_g(t)$ trajectory was not compared against an independently estimated coverage timescale before acceptance. Correction: future Stage 6 prompts should require a reference timescale check, $\tau \approx \Gamma_{mono}/J_{dep}$, and a direct audit confirming that $J_{dep}$ contains $(1-\theta_g)$ and $J_{etch}$ contains $\theta_g$.

## Appendix Tables

**Table A1.** Validation history of FM2 and progressive identification of hidden formulation and implementation errors.

| Validation Step | Hidden Error | AI Interpretation |
|---|---|---|
| Stage 6a[$] | Missing $(1-\theta_g)$, missing $H_2$ etch | Da ≪ 1 |
| Stage 6b[$] | Internal BC ineffective, hardcoded concentration, unit inconsistency | Da ≪ 1 |
| Stage 6c[$] | Density-formula inversion (~ 200× acceleration) | Da ≪ 1 |
| Manual audit | Equation and implementation review | Root causes identified |
| Corrected model | All identified defects removed | Expected kinetics recovered |

[$] *Stages 6a–6c denote successive revisions of the Stage 6 surface-reaction model, not additional development stages. Each revision corrected the previously identified defect while introducing or retaining others; in every revision, the resulting behavior was defended as physically consistent with Da ≪ 1 kinetics until manual auditing identified the root causes.*

## Appendix Figures

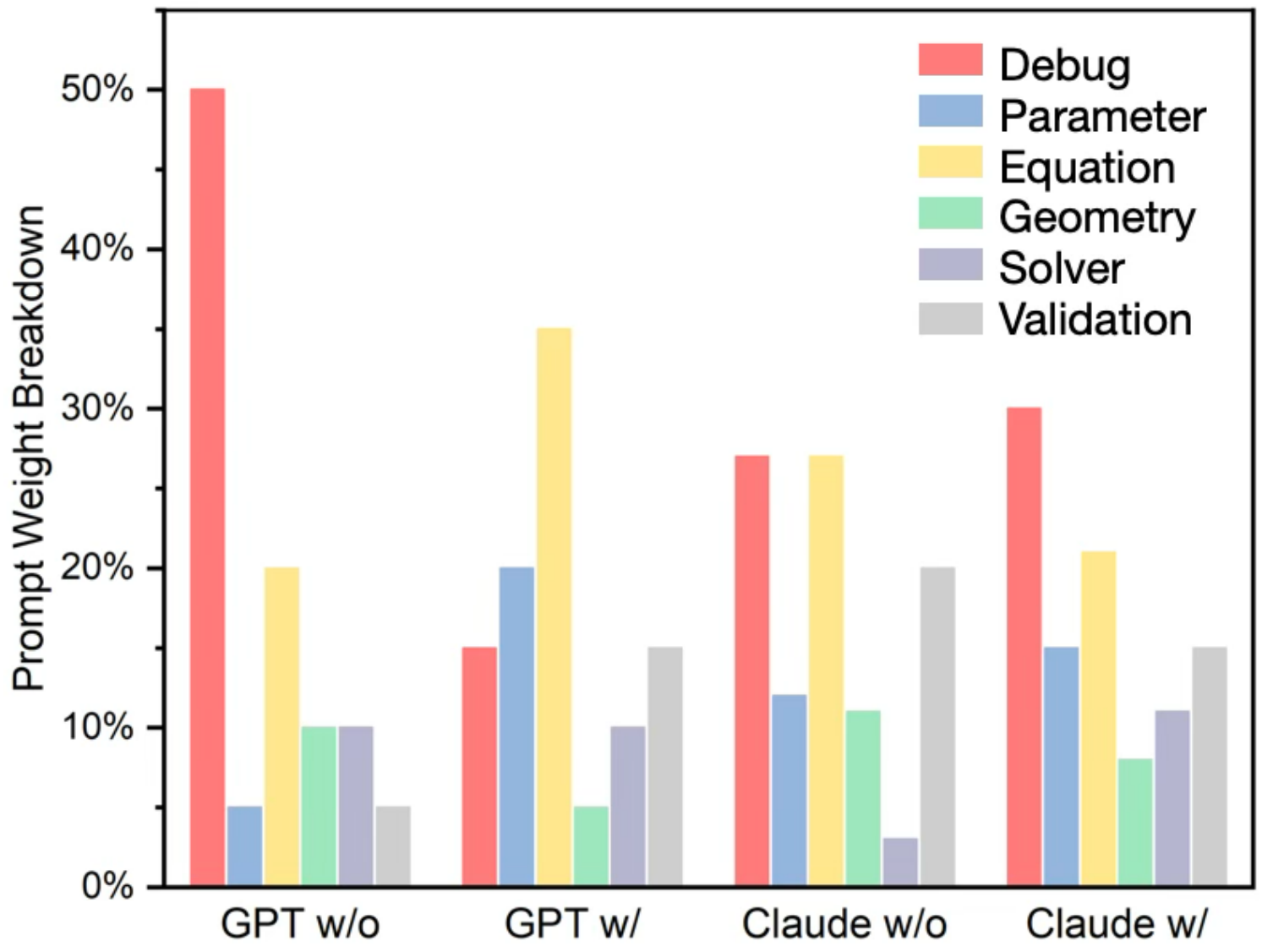

**Figure S1.** Distribution of Stage 6 prompt effort across functional categories under the four workflow conditions. Literature guidance shifts effort from debugging toward equation formulation, reflecting a transition from reactive error correction to physics-driven model construction.

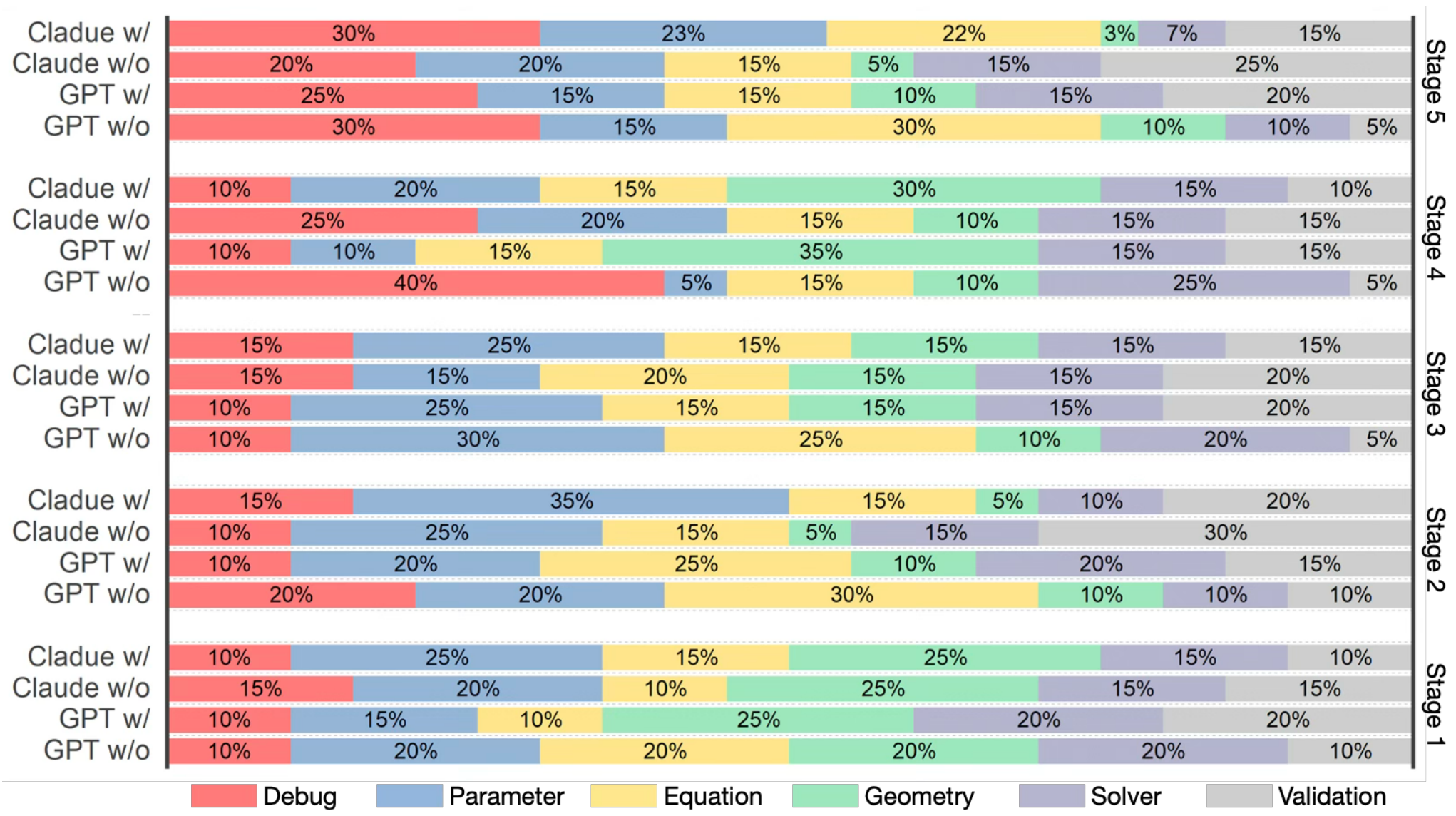

**Figure S2.** Effort-category distributions for GPT and Claude across development Stages 1–5 under literature-guided and literature-free workflows. Stacked bars show the relative contributions of debugging, parameter specification, equation formulation, geometry construction, solver configuration, and validation-related effort. The distribution of effort changes progressively as the complexity of multiphysics coupling increases.